\documentclass[12pt,showpacs]{revtex4}

\usepackage{amssymb}
\usepackage{amsmath}
\usepackage{graphicx}

\newcommand{\bq}{\begin{eqnarray}}
\newcommand{\eq}{\end{eqnarray}}
\newcommand{\bqn}{\begin{eqnarray*}}
\newcommand{\eqn}{\end{eqnarray*}}

\begin{document}
\title{ Stability of the iterative solutions of integral equations as one
phase freezing criterion}

\author{R. Fantoni}
\email{rfantoni@ts.infn.it}
\affiliation{Dipartimento di Fisica Teorica dell' Universit\`a  
and Istituto Nazionale di Fisica della Materia, Strada Costiera 11, 
34014 Trieste, Italy}
\author{G. Pastore}
\email{pastore@ts.infn.it} 
\affiliation{Dipartimento di Fisica Teorica dell' Universit\`a  
and Istituto Nazionale di Fisica della Materia, Strada Costiera 11, 
34014 Trieste, Italy}
\date{\today}

\begin{abstract}
\noindent
A recently proposed connection between the threshold for the stability of 
the iterative solution of integral equations for the pair correlation 
functions of a classical fluid and the structural instability of the 
corresponding real fluid is carefully analyzed.
Direct calculation of the Lyapunov exponent of the standard iterative 
solution of HNC and PY integral equations for the 1D hard rods fluid shows 
the same behavior observed in 3D systems. Since no phase transition is 
allowed  in such 1D system, our analysis shows that the proposed one 
phase criterion, at least in this case, fails. We argue 
that the observed proximity between the numerical and the 
structural instability in 3D originates from the enhanced structure  
present in the fluid but, in view of the arbitrary dependence on the 
iteration scheme, it seems uneasy  to relate the  numerical stability 
analysis to a robust one-phase criterion for predicting a thermodynamic 
phase transition.
\end{abstract}

\pacs{05.70.Fh, 61.20.Ne}

\maketitle

\section{Introduction}

When studying the structure and thermodynamics of classical fluids one is often
faced with the task of solving the nonlinear integral equation which
stems out of the combination of the Ornstein-Zernike equation and
an approximate relation between pair potential and correlation functions (the 
closure) \cite{Hansen}. Integral equations can be
generally written in the form
\bq \label{intro:ie}
\gamma(r)=A\gamma(r)~~~,
\eq
where $\gamma(r)\in S$ may be the total correlation function $h(r)$,
the direct correlation function $c(r)$, or a combination of the two,
$S$ is a set of a metric space of functions, and $A:S\rightarrow S$ is a 
non linear operator mapping $S$ into itself.

Numerical analysis of integral equations suggests the use of the
following combination 
\bq \label{icf}
\gamma(r)=h(r)-c(r)~~~,
\eq
since $\gamma$ is a much smoother function than $h$ or $c$, especially
in the core region.

It has been pointed out by Malescio {\sl et. al.}
\cite{Malescio98,Malescio00a,Malescio00b} that, amongst the different
numerical schemes that one may choose to solve (\ref{intro:ie}), the
{\sl simple iterative scheme of Picard} plays a special role. Picard
scheme consists in generating successive approximations to the
solution through the relationship 
\bq \label{picard}
\gamma_{n+1}=A\gamma_n~~~,
\eq
starting from some initial value $\gamma_0$. If the sequence of
successive approximations $\{\gamma_n\}$ converges toward a value
$\gamma^\star$, then $\gamma^\star$ is a fixed point for the operator $A$,
i.e. it is a solution of Eq. (\ref{intro:ie}),
$\gamma^\star=A\gamma^\star$. 
Banach' s fixed point theorem (see chapter 1 in
\cite{Zeidler} especially theorem 1.A) states that, given an
operator $A:S\rightarrow S$, where $S$ is a closed nonempty set in a
complete metric space, the simple iteration (\ref{picard}) may converge 
toward the only fixed point in $S$ ($A$ is $k$ contractive) or it may 
not converge ($A$ is non expansive). So the simple iterative method
can be used to signal a fundamental change in the properties of
the underlying operator.

The operator $A$ will in general depend on the thermodynamic 
state of the fluid. In order to determine the properties of the
operator at a given state we can proceed as follows. First, we 
find the fixed point $\gamma^\star$ using a numerical scheme (more refined
then Picard' s) capable of converging in the high density region. Next,
we perturb the fixed point with an arbitrary initial perturbation
$\delta_0(r)$ so that 
\bq
A(\gamma^\star+\delta_0)\simeq A\gamma^\star+\left.\frac{\partial
A}{\partial\gamma}\right|_{\gamma^\star}\delta_0=\gamma^\star+M\delta_0~~~,
\eq
where we have introduced the Floquet matrix $M$. Now
$\delta_1=M\delta_0$ may be considered as the new perturbation. We
then generate the succession $\{\delta_n\}$ where
\bq
\delta_n=M\delta_{n-1}~~~.
\eq
If the succession converges to zero then the operator $A$ is $k$
contractive, if it diverges the operator is non expansive. Malescio
{\sl et. al.} call $\{\delta_n\}$ {\sl fictitious dynamics} and
associate to the resulting fate of the initial perturbation the nature
of the {\sl structural equilibrium} of the fluid. If the succession
converges to zero they say that the fluid is {\sl structurally
stable} and {\sl structurally unstable} otherwise. We will call
$\rho_{inst}$ the density where the transition between a structurally 
stable and unstable fluid occurs.

Following Malescio {\sl et. al.} it is possible to define a {\sl
measure} for the structural stability of the system as follows. We
define  
\bq
S_i=\frac{||M\delta_i(r)||}{||\delta_i(r)||}~~~,
\eq
where $||f(r)||=\sqrt{\sum_{i=1}^Nf^2(r_i)}$ is the norm of a function
$f$ defined over a mesh of $N$ points. We assume that the norm of the
perturbation depends exponentially on the number of iterations 
\bq
||\delta_n||=||\delta_0||2^{\lambda n}~~~,
\eq
where $\lambda$ is the Lyapunov exponent related to the fictitious
dynamics. Then one can write the average exponential stretching of
initially nearby points as
\bq \label{intro:lya1}
\lambda=\lim_{n\rightarrow\infty}\frac{1}{n}\log_2\left(\prod_{i=0}^{n-1}S_i
\right)~~~.
\eq

Malescio {\sl et. al.} have calculated the dependence of $\lambda$ on
the density for various simple three dimensional liquids (and various
closures): hard spheres \cite{Malescio98}, Yukawa, inverse
power and Lennard-Jones potentials \cite{Malescio00a}. For
all these systems they found that $\lambda$ increases with the density 
and the density at which $\lambda$ becomes positive,
$\rho_{inst}$, falls close to the 
freezing density $\rho_f$ of the fluid system. This
occurrence lead them to propose  this kind of
analysis as a one-phase criterion to predict the freezing
transition of a dense fluid and to estimate $\rho_f$.
However, we think that there are some practical and conceptual 
difficulties with 
such one-phase criterion. 

First of all,   it does not depend only
on the closure adopted but  also on the
kind of algorithm used to solve the integral equation. Indeed, different
algorithms give different $\rho_{inst}$ and Malescio {\sl et. al.}
choose to use as instability threshold for their criterion the one
obtained using Picard algorithm, thus giving to it a special status. 
However, it is hard to understand why the 
particular algorithm adopted in the solution of the
integral equation should be directly related to a phase boundary.

Moreover, one would expect that the estimate of $\rho_{inst}$ would improve in
connection with improved closures. This is not the case, at least in the 
one component
hard sphere fluid.

Even a more serious doubt about the validity of the proposed criterion 
comes from 
its behavior in one dimensional systems. 
In this paper we present the same Lyapunov exponent
analysis on a system of hard rods in one dimension treated using
either the Percus-Yevick (PY) or the hypernetted chain (HNC)
approximations. What we find is that the Lyapunov exponent as a
function of density has the same behavior as that for the three
dimensional system (hard spheres): it becomes positive beyond a
certain $\rho_{inst}$. Since it is known \cite{VanHove50} that a one
dimensional fluid of hard rods does not have a phase transition,
our result  sheds some doubts on the validity of the proposed 
criterion. 

\section{Technical details}

As  numerical scheme to calculate the fixed point we
used Zerah' s algorithm \cite{Zerah85} for the three dimensional
systems and a modified iterative method for the hard rods in one
dimension. In the modified iterative method input and output are mixed
at each iteration
\bq \label{mis}
\gamma_{n+1}=A_{mix}\gamma_n=\alpha A\gamma_n+(1-\alpha)\gamma_n~~~, 
\eq
where $\alpha$ is a real parameter $0<\alpha<1$. Note that while for a
non expansive operator $A$ the Picard iterative method (\ref{picard})
needs not converge, one can prove convergence results on an Hilbert
space for the modified iterative method with fixed $\alpha$ (see
proposition 10.16 in \cite{Zeidler}).  
In all the computations we used a uniform grid of $N=1024$ points with a 
spacing $\delta r=0.025$. Generally, we observed a marginal increase of
$\rho_{inst}$ by lowering $N$. 

A method to find a Lyapunov exponent, equivalent but more accurate than 
the one of
Malescio {\sl et. al.} (\ref{intro:lya1}), goes through the
diagonalization of the Floquet matrix. Note that in general this
matrix is non symmetric, thus yielding complex eigenvalues. A Lyapunov
exponent can then be defined as \cite{Sagdeev}
\bq \label{intro:lya2}
\lambda^\prime=\log\left[\max_i\left(\sqrt{er_i^2+ei_i^2}\right)\right]
~~~,  
\eq 
where $er_i$ and $ei_i$ are respectively the real and imaginary part
of the $i$-th eigenvalue. 
In our numerical computations we always used recipe (\ref{intro:lya2})
to calculate the Lyapunov exponents since it is explicitly independent
from the choice of an initial perturbation.

We constructed the Floquet matrix in the following way \cite{Gillan79}.
In a Picard iteration we start from $\gamma(r)$ we calculate $c(r)$
from the closure approximation, we calculate its Fourier transform
$\tilde{c}(k)$, we calculate $\tilde{\gamma}(k)$ from the OZ equation,
and finally we anti transform $\tilde{\gamma}$ to get
$\gamma^\prime(r)$. For example for a three dimensional system a PY
iteration in discrete form can be written as follows 
\bq
c_i&=&(1+\gamma_i)\left(e^{-\beta\phi_i}-1\right)~~~,\\\label{ck}
\tilde{c}_j&=&\frac{4\pi\delta r}{k_j}\sum_{i=1}^{N-1}r_i\sin(k_jr_i)
c_i~~~,\\
\tilde{\gamma}_j&=&\rho\tilde{c}_j^2/(1-\rho\tilde{c}_j)~~~\\\label{gr}
\gamma_i^\prime&=&\frac{\delta k}{2\pi^2r_i}\sum_{j=1}^{N-1}k_j
\sin(k_jr_i)\tilde{\gamma}_j~~~, 
\eq
where $r_i=i\delta r$ are the $N$ mesh points in $r$ space,
$k_j=j\delta k$ are the $N$ mesh points in $k$ space, with $\delta
k=\pi/(N\delta r)$, $c_i=c(r_i)$, $\gamma_i=\gamma(r_i)$,
$\tilde{c}_j=\tilde{c}(k_j)$, $\tilde{\gamma}_j=\tilde{\gamma}(k_j)$,
and $\phi_i=\phi(r_i)$ is the interparticle potential calculated on
the grid points. The Floquet matrix will then be
\bq \nonumber
M_{ij}&=&\frac{\partial\gamma_i^\prime}{\partial\gamma_j}=
\sum_{m=1}^{N-1}\frac{\partial\gamma_i^\prime}{\partial \tilde{\gamma}_m}
\frac{\partial \tilde{\gamma}_m}{\partial\tilde{c}_m}
\frac{\partial\tilde{c}_m}{\partial c_j}
\frac{\partial c_j}{\partial\gamma_j}\\ \label{floquet}
&=&\frac{\delta r\delta k}{\pi}\left(\frac{r_j}{r_i}\right)\left(
e^{-\beta\phi_j}-1\right)(D_{i-j}-D_{i+j})~~~,
\eq
where
\bq
D_l=\sum_{m=1}^{N-1}\cos(k_mr_l)\left[
\frac{2\rho\tilde{c}_m}{1-\rho\tilde{c}_m}+
\left(\frac{\rho\tilde{c}_m}{1-\rho\tilde{c}_m}\right)^2\right]~~~.
\eq

The HNC case can be obtained replacing in
(\ref{floquet}) $[\exp(-\beta\phi_j)-1]$ with
$[\exp(-\beta\phi_j+\gamma_j)-1]$. 

To derive the expression for the Floquet matrix valid for the one
dimensional system and consistent with a trapezoidal discretization of
the integrals, we need to replace (\ref{ck}) and (\ref{gr}) with
\bq
\tilde{c}_j&=&2\delta r\left(\sum_{i=1}^{N-1}\cos(k_jr_i)c_i+\frac{1}{2}
c_0\right)~~~,\\
\gamma^\prime_i&=&\frac{\delta k}{\pi}\left(\sum_{i=1}^{N-1}
\cos(k_jr_i)\tilde{\gamma}_j+\frac{1}{2}\tilde{\gamma}_0\right)~~~.
\eq
\section{Numerical results}

We checked our procedure  for a three dimensional hard
spheres fluid and a Lennard-Jones fluid at a reduced temperature
$T^*=2.74$. Our results, obtained using recipe (\ref{intro:lya2}), 
were in good agreement with those of Malescio {\sl et. al.} 
\cite{Malescio98,Malescio00a} which used recipe
(\ref{intro:lya1}) instead (another difference between our analysis
and theirs is that we used for $\gamma$ the indirect correlation
function (\ref{icf}) while they were using the total correlation 
function $h$). For the Lennard-Jones fluid our results where indistinguishable 
from those
of Malescio {\sl et. al.} \cite{Malescio00a}. We found a reduced
instability density $\rho^*_{inst}$ around 1.09 in the PY
approximation and around 1.06 in the HNC approximation. 
For the three dimensional hard sphere
fluid we found slightly larger ($4\%$)  values for $\rho_{inst}$. 
We found a
$\eta_{inst}=\rho_{inst}\pi d^3/6$ of about 0.445 in the PY
approximation and around 0.461 in the HNC approximation.  We also checked the
value corresponding to the 
Martynov-Sarkisov (MS) \cite{MS} closure and we found
$\eta_{inst}= 0.543$. 

We feel that the differences are within what we can 
expect on the basis of small numerical differences in different 
implementations.
We think that it is more worth of notice that closures providing better
structural and thermodynamic properties, like PY or MS do not provide a 
better value of $\eta_{inst}$.

We looked at the structure of the Floquet matrix too but from direct
inspection we can conclude that it is not diagonally dominated.

Then, we have calculated the Lyapunov exponent (\ref{intro:lya2}) as a
function of the density for a fluid of hard rods in one dimension 
using both PY and HNC closures. The results of the calculation are
shown in Fig. \ref{fig:le1d-1} and Fig. \ref{fig:le1d-2}. The curves show
the same qualitative behavior as the ones for the three dimensional fluid.
From Fig. \ref{fig:le1d-1} we can see how the slope of the curves
starts high at low densities and decreases rapidly with $\rho$. At
high densities the Lyapunov exponent becomes zero at $\rho_{inst}$.
As expected, to find the fixed point for $\rho\gtrsim\rho_{inst}$ it
is necessary to choose $\alpha<1$ in the modified iterative scheme
(\ref{mis}). 
Before reaching the instability threshold the curves show a rapid
change in their slope at $\rho_c<\rho_{inst}$. Figure \ref{fig:le1d-2}
shows a magnification of the region around $\rho_c$ from which
we are lead to conclude that, within the numerical accuracy of the
calculations, the slope of the curves $d\lambda^\prime/d\rho$
undergoes a jump at $\rho_c$.

\section{Conclusions}

The fictitious dynamics associated to the iterative solution of an
integral equation can signal the transition of the map of the
integral equation from $k$ contractive to non expansive. If the
Lyapunov exponent is negative the map is $k$ contractive, if it is
positive the map is non expansive. 

Since it is possible to modify in an arbitrary way the fictitious dynamics 
keeping the 
same fixed point, it is difficult to understand  a deep 
direct connection between the stability properties of the map and a one-phase 
criterion for a thermodynamic transition.

Admittedly the correlations shown by Malescio et al. are striking.
We calculated the Lyapunov exponent as a function of the density for
various fluids (hard spheres in one and three dimensions and three
dimensional Lennard-Jones fluid) both in the HNC and PY
approximations. 
For the three dimensional fluids the instability density falls close to
the freezing density $\rho_f$. For example, the Lennard-Jones fluid 
studied with HNC should undergo a
freezing transition at $\rho^*\simeq 1.06$ or at
$\rho^*\simeq 1.09$, if studied with PY ,  rather close to the 
freezing density $\rho_f^*\simeq 1.113$.
For hard spheres
$\rho_{inst}^*$ is about $10\%$ smaller than $\rho_f^*\sim 0.948$.
The Hansen-Verlet ``rule'' states that a simple fluid freezes when the
maximum of the structure factor is about 2.85
\cite{Hansen69}. According to this rule the three dimensional hard
spheres fluid studied
with HNC should undergo a freezing transition at $\rho\simeq 1.01$
while when studied with PY the transition should be at $\rho\simeq
0.936$. The corresponding estimates obtained through $\rho_{inst}^*$,
$0.879$ (HNC) and $0.850$ (PY) are poorer and, more important, are not
consistent with the well known better performance of PY in the case of
hard spheres. 

\begin{figure}[hbt]
\begin{center}
\includegraphics[width=10cm]{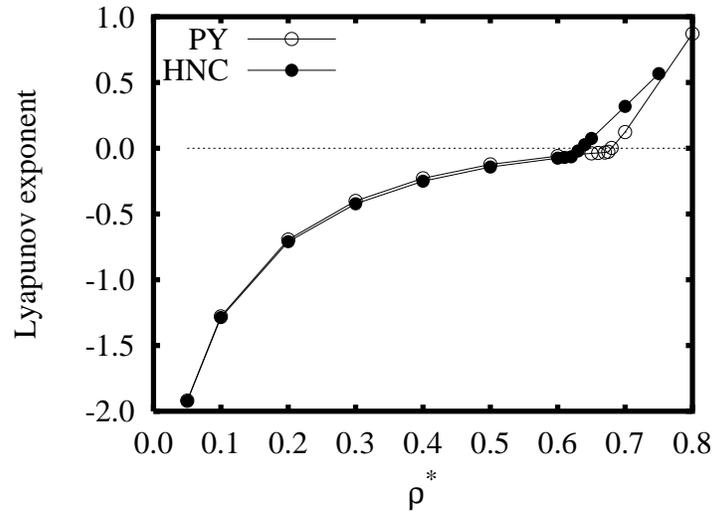}
\end{center}
\caption[For a fluid of hard rods in one dimension, we show the
Lyapunov exponent as a function of the reduced density
($\rho^*=\rho\sigma$ where $\sigma$ is the rods width) as 
calculated using the PY and the HNC closures.]{R. Fantoni and G. Pastore 
\label{fig:le1d-1}
}
\end{figure}

\begin{figure}[hbt]
\begin{center}
\includegraphics[width=10cm]{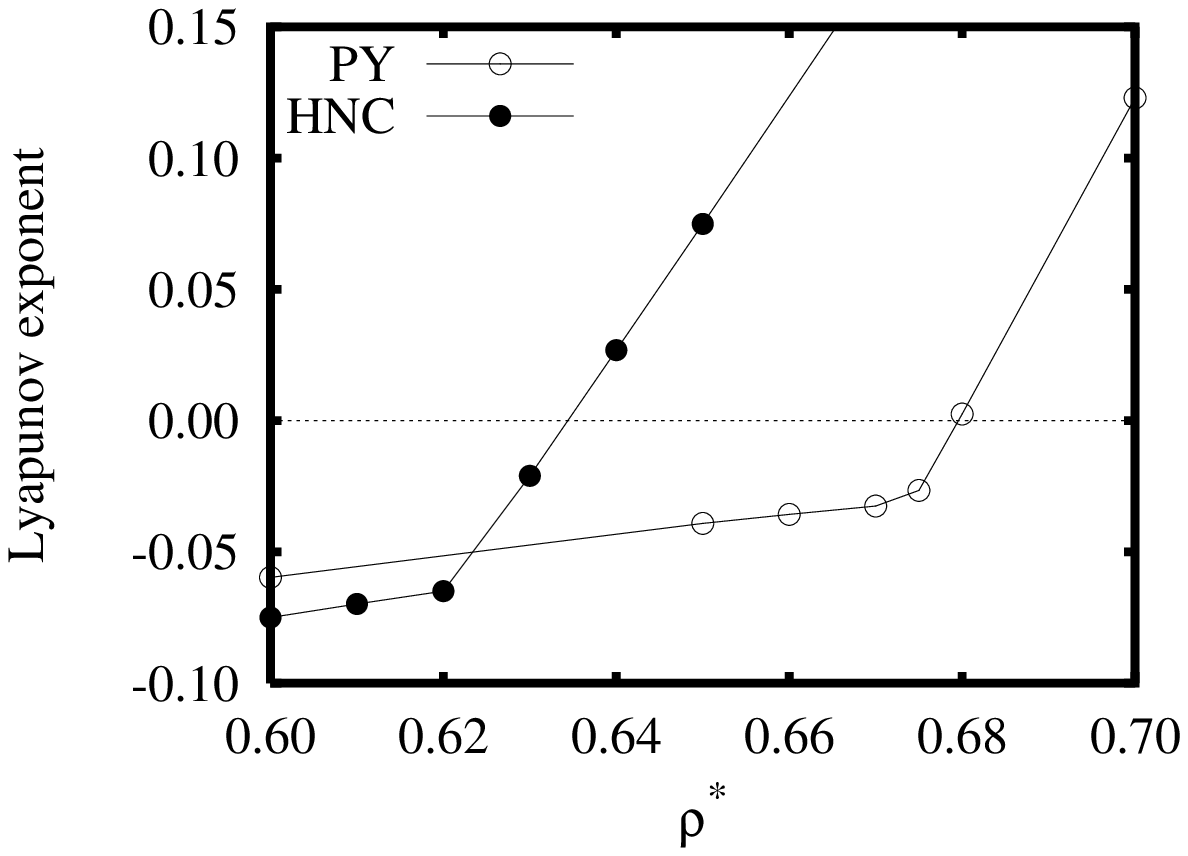}
\end{center}
\caption[We show a magnification of Fig. \ref{fig:le1d-1} in a
neighborhood of the instability threshold.]{R. Fantoni and G. Pastore
\label{fig:le1d-2}
}
\end{figure}

In one dimension, a fluid of hard spheres (hard rods), cannot undergo a
phase transition \cite{VanHove50}. From Fig. \ref{fig:le1d-1} 
we see that the system still becomes structurally
unstable. This can be explained by observing that the structural
stability as defined by Malescio {\sl et. al.} is a property of the 
map $A$ and in particular of the algorithm used to get solution of 
the integral equation under study.  
As such, it is not directly related to the  thermodynamic
properties even at the approximate level of the theory (there is no
direct relation between the contractiveness properties of $A$ and the
thermodynamics). It looks more reasonable
that the increase of the correlations would be the common origin 
of the numerical instability of the Picard iteration 
and, whenever it is possible, of thermodynamic phase transitions.

\begin{acknowledgments}

G.P. would like to acknowledge preliminary exploratory work on this subject 
carried on in collaboration with Matteo Mosangini and Waheed Adeniyi Adeagbo.

\end{acknowledgments}
\bibliography{stabiter}

\newpage
\centerline{\bf LIST OF FIGURES}
\begin{itemize}
\item[Fig. 1] For a fluid of hard rods in one dimension, we show the
Lyapunov exponent as a function of the reduced density
($\rho^*=\rho\sigma$ where $\sigma$ is the rods width) as 
calculated using the PY and the HNC closures.
\item[Fig. 2] We show a magnification of Fig. \ref{fig:le1d-1} in a
neighborhood of the instability threshold.
\end{itemize}

\newpage
\newpage
%
\end{document}